\documentclass[12pt,a4paper,final]{iopart}

\newcommand{\be}{\begin{equation}}
\newcommand{\ee}{\end{equation}}
\newcommand{\ba}{\begin{eqnarray}}
\newcommand{\ea}{\end{eqnarray}}
\newcommand{\beq}{\begin{equation}}
\newcommand{\eeq}{\end{equation}}
\newcommand{\beqa}{\begin{eqnarray}}
\newcommand{\eeqa}{\end{eqnarray}}

\usepackage{iopams}
\usepackage{graphicx}
\usepackage{amsmath}
\usepackage{subfigure}
\usepackage[breaklinks=true,colorlinks=true,linkcolor=blue,urlcolor=blue,citecolor=blue]{hyperref}
\usepackage[normalem]{ulem}
\usepackage[labelfont=bf,font=small]{caption}
\begin{document}

\title[Snapping swallowtails in accelerating black hole thermodynamics]{Snapping Swallowtails in Accelerating Black Hole Thermodynamics}

\author{Niloofar Abbasvandi}
\address{Department of Physics and Astronomy, University of Waterloo, Waterloo,
Ontario, Canada, N2L 3G1}
\ead{niloofar.abbasvandi@uwaterloo.ca}

\author{Wan Cong}
\address{Department of Physics and Astronomy, University of Waterloo, Waterloo,
Ontario, Canada, N2L 3G1}
\address{Perimeter Institute, 31 Caroline St., Waterloo, Ontario, N2L 2Y5, Canada}
\ead{wcong@uwaterloo.ca}

\author{David Kubiz\v n\'ak}
\address{Perimeter Institute, 31 Caroline St., Waterloo, Ontario, N2L 2Y5, Canada}
\address{Department of Physics and Astronomy, University of Waterloo, Waterloo,
Ontario, Canada, N2L 3G1}
\ead{dkubiznak@perimeterinstitute.ca}

\author{Robert B. Mann}
\address{Department of Physics and Astronomy, University of Waterloo, Waterloo,
Ontario, Canada, N2L 3G1}
\address{Perimeter Institute, 31 Caroline St., Waterloo, Ontario, N2L 2Y5, Canada}
\ead{rbmann@uwaterloo.ca}

\begin{abstract}
The thermodynamic behaviour of a charged and accelerating AdS black hole is studied in the context of extended phase space with variable cosmological constant. When compared to the charged AdS black hole  without acceleration, a remarkable new feature of `snapping swallowtails' appears. Namely, for a black hole with any charge $Q$  and any string tension $\mu$ causing the acceleration of the black hole, there exists a transition pressure $P_t=3\mu^2/(8\pi Q^2)$ at which the standard swallowtail `snaps', causing the branch of low temperature black holes to completely disappear, leading to a pressure induced zeroth order phase transition between small and large black holes. For intermediate values of the string tension, we also observe a reentrant phase transition, as the small black hole changes to a large one and then back to small, as the pressure decreases, crossing the coexistence line of the two phases several times. We also find a new class of `mini-entropic' black holes, whose isoperimetric ratio becomes unbounded in a certain region of parameter space.

\end{abstract}
\pagebreak
\section{Introduction}
The discovery that black holes can be assigned thermodynamic temperature \cite{Hawking:1974sw} and entropy \cite{Bekenstein:1973ur, Bekenstein:1974ax} inaugurated the era of black hole thermodynamics.
 Featuring   radiation that is universal over all particle species, an entropy that is proportional to the black hole area rather than volume,  and negative specific heat,   black hole thermodynamics has continued to be full of surprises.
Especially interesting are the
 thermodynamic properties of black holes in asymptotically AdS spacetimes. These objects are known to exhibit interesting phase behaviour \cite{Hawking:1982dh, Chamblin:1999tk,Chamblin:1999hg}, relevant for the description of strongly coupled CFTs via the AdS/CFT correspondence \cite{Witten:1998zw}.

Moreover, in  AdS black hole spacetimes one can consider the negative cosmological constant  $\Lambda$ to be
a thermodynamic quantity whose variation gives rise to an additional pressure--volume term in the first law of black hole  thermodynamics \cite{CreightonMann:1995, Caldarelli:1999xj, Kastor:2009wy, Dolan:2010ha, Cvetic:2010jb, Dolan:2011xt, Kubiznak:2012wp}. This is done via the identification
\begin{equation}
    P = -\frac{\Lambda}{8 \pi}\,,
\end{equation}
 relating  $\Lambda$ to the thermodynamic pressure $P$.
 The subsequent study of black hole thermodynamics in this {\em extended phase space} has revealed   remarkable similarities between the phase behaviour of black holes and that of ordinary matter \cite{Kubiznak:2016qmn}.

\begin{figure}[t]
    \centering
    \includegraphics[scale=0.65]{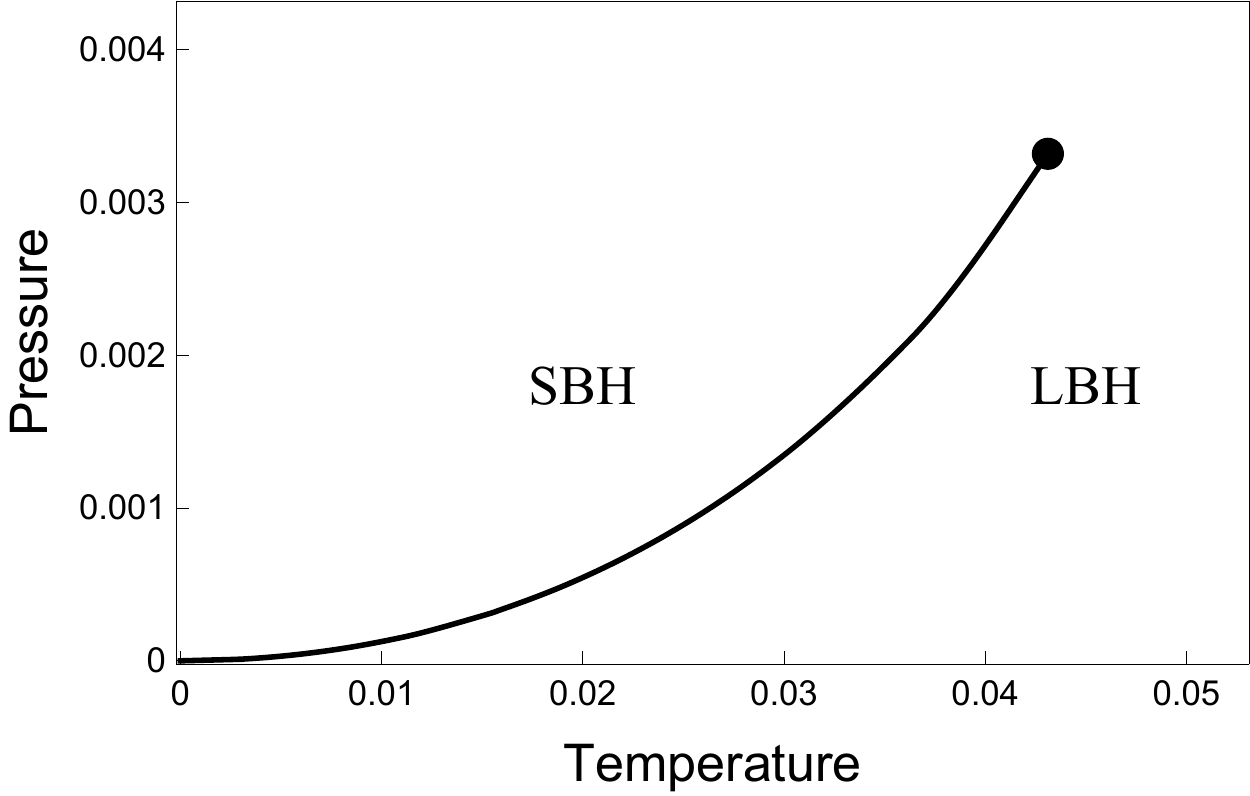}
    \caption{\textbf{Phase diagram of a charged AdS black hole.} The $P-T$ diagram of a charged AdS black hole without acceleration is reminiscent of what happens for the liquid/gas phase transition. The coexistence line between small black hole (SBH) phase (analogous to the liquid phase) and large black hole (LBH) phase (analogous to the gas phase) emerges from the origin on one end and terminates at a critical point at $(T_c^{(0)}, P_{c}^{(0)})$ on the other end. The plot is made for $Q=1$. }
    \label{fig1}
\end{figure}

The paradigmatic example  of this analogy is the Van der Waals-like phase transition
of a charged AdS black hole
\cite{Chamblin:1999tk,Chamblin:1999hg, Kubiznak:2012wp}. In this case, the first law takes the following form:
\begin{equation}
\label{eq: RN first law}
    \delta M  = T\delta S+\Phi \delta Q+V\delta P\,,
\end{equation}
where $V$ stands for the thermodynamic volume of the black hole, a quantity conjugate to the thermodynamic pressure $P$, and $M\,,\, T\,,\, S\,,\, \Phi\,,$ and  $Q$ are respectively  the black hole  mass, temperature, entropy, electric potential, and charge.
In a canonical  (fixed charge)  ensemble the system exhibits a phase transition from a small black hole to large black hole, fully analogous to the liquid/gas phase transition of the Van der Waals fluid \cite{Kubiznak:2012wp}.  Namely, for pressures below a critical pressure, $P<P_c^{(0)}$, the free energy of the system exhibits  swallowtail behaviour, characteristic of the first order phase transition. The two black hole phases  that meet at the `bottom of the swallowtail' discontinuously differ in horizon radius. The corresponding phase diagram (the $P-T$  diagram) reproduced in Fig.~\ref{fig1} exhibits a coexistence line, which emanates from the origin and terminates at a (standard) {\em critical point}, characterized by the critical temperature and pressure
\be
T_c^{(0)} =\frac{\sqrt{6}}{18 \pi Q}\,,\quad P_c^{(0)}=\frac{1}{96\pi Q^2}\,.
\ee
 At this critical point the phase transition becomes of the second order and is characterized by the Ising universality class mean field theory critical exponents \cite{Kubiznak:2012wp}.

Remarkably,  Van der Waals-like behaviour has since been observed for many rotating and/or charged AdS black holes in various spacetime dimensions and for various gravity theories. Although more complicated phase behaviour may occur (see \cite{Kubiznak:2016qmn} for a review),  Van der Waals behaviour can be regarded as prototypical thermodynamic behaviour for AdS black holes in a canonical ensemble. For this reason it is rather  instructive to find a departure from this prototypical behaviour and seek more complicated phase diagrams for AdS black holes,  reflecting thus a more complicated behaviour of the dual CFT. To this end the following features have been identified: reentrant phase transitions \cite{Altamirano:2013ane},  triple points analogous to that of water \cite{Altamirano:2013uqa}, isolated critical points \cite{Dolan:2014vba,Frassino:2014pha}, and superfluid-like behaviour \cite{Hennigar:2016xwd}.

Motivated by the above, we investigate the thermodynamic behaviour of charged accelerating AdS black holes.
The history of accelerating black holes goes all the way back to the early days of general relativity to the discovery of the  {\em C-metric} \cite{Weyl:1917gp,levi1918ds2}.  This metric has  played an important role for many developments in general relativity, has been rediscovered many times (see \cite{kinnersley1970uniformly} for references), and is understood now to describe accelerating black holes.
The C-metric yields an example of an exact radiative spacetime and was exploited to study radiative patterns in spacetimes with various asymptotics e.g. \cite{Podolsky:2003gm}. It has been  used to study  black hole nucleation in various backgrounds \cite{Dowker:1993bt, Ross:2002dr},
to provide the means for splitting  a cosmic string \cite{Gregory:1995hd, Eardley:1995au}, and in a generalized form   was  used to construct 5-dimensional black rings \cite{Emparan:2001wn}. However its thermodynamic behaviour remains largely unexplored, and it is our purpose in this paper to begin to rectify this situation.

We shall restrict our attention to  charged {\em slowly} accelerating AdS black holes. The associated C-metric  is a special case of the general Plebanski--Demianski class of solutions of the Einstein---Maxwell equations \cite{Plebanski:1976gy}. It describes a single accelerated black hole in AdS space that is suspended on a ``cosmic string'' represented by a conical deficit on  one of the polar axes.   Since
there is only one horizon (no acceleration or cosmological horizons are present in the limit of slow acceleration), the system has a unique temperature and its thermodynamics can be defined   \cite{Anabalon:2018ydc,Anabalon:2018qfv} (see also \cite{Appels:2016uha,Appels:2017xoe,Gregory:2017ogk, Astorino:2016xiy,
Astorino:2016ybm, Zhang:2018hms} for  prior investigations). The key observation is that the tension $\mu$  of the string, that causes the black hole to accelerate, can be treated as (yet another) thermodynamic quantity whose variations add a new  work term to the first law ~\eqref{eq: RN first law}, which now reads
\be\label{two}
\delta M  = T\delta S+\Phi \delta Q+V\delta P-\lambda \delta \mu\,,
\ee
where $\lambda$ is a conjugate thermodynamic quantity to $\mu$, known as the thermodynamic length \cite{Appels:2017xoe}.


As we shall see, the presence of acceleration  has far-reaching consequences for the phase behaviour of charged AdS black holes.
Let us first consider tiny acceleration (small string tension). In this case we find that the high pressure phase behaviour remains qualitatively similar to that of the non-accelerating charged black hole (with swallowtails and a critical point at $P_c\approx P_c^{(0)}$ above). However, a fascinating novel feature occurs at low pressures: for any charge $Q$ and tension $\mu$, there exists a critical pressure
\be\label{Pt1}
P_t=\frac{3\mu^2}{8\pi Q^2}\,,
\ee
at which the {\em swallowtail  `snaps'}. As the pressure decreases below $P_t$, the entire branch of small stable low temperature black holes disappears, breaking the structure of the swallowtail, and a new branch of unstable high temperature black holes emerges. This corresponds to a
breaking of the co-existence line in the $P-T$ plane, which now no longer continues all the way to the origin but rather terminates at a new ``{critical point}'', at $(T_t, P_t)$,  c.f. Fig.~\ref{fig1} with Fig.~\ref{fig: RPT}.  Since the lower branch of the swallowtail is absent, a ``no black hole'' region at low temperatures emerges for $T<T_0$. Moreover, there is a small range of temperatures, between $T_0$ and $T_t$, where the global minimum of the free energy ``jumps'' as it crosses $P_t$, corresponding to a zeroth order phase transition from small to intermediate black holes. Since the lines of coexistence of the first
and zeroth order phase transitions merge at the point $(T_t, P_t)$ (as we shall illustrate below)
 we shall refer to this point as  a ``{\em bicritical point''}.

For small tensions we have $P_t\ll P_c$.
As the tension of the string increases, the bicritical and the (standard) critical point move closer to each other, with the former moving towards higher temperatures and pressures and the latter towards lower temperatures and pressures. At the same time the coexistence line of the first order phase transition exhibits a change in convexity and starts tilting towards higher temperatures as the pressure decreases. This gives rise to a {\em reentrant phase transition} \cite{Altamirano:2013ane}, with pressure being the thermodynamic quantity that is monotonically varied, yielding a small  to large  and back to small black hole phase transition (as we shall later illustrate). Finally for sufficiently large tensions, $\mu\approx 1/4$, the slope of the coexistence line is always negative  and does not resemble the behaviour of the liquid/gas phase transition any longer.

We also find a rather interesting limiting point in thermodynamic phase space at which
the volume diverges but the entropy remains finite.   We refer to black holes whose parameters are in the neighbourhood of
this point as {\em mini-entropic}, since their  dimensionless volume-to-area (or isoperimetric) ratio \cite{Cvetic:2010jb}
 \be\label{eq:ipe-ratio}
\mathcal{R} = \left(\frac{(d-1) {V}}{\Omega_{d-2}}\right)^{\frac{1}{d-1}}\left(\frac{\Omega_{d-2}}{ {{\mathcal A}}}\right)^{\frac{1}{d-2}}
\ee
becomes unbounded as this point is approached,  contrary to the superentropic black holes \cite{Hennigar:2014cfa} that are characterized by a ratio less than one.  In the above equation, $V$ is the thermodynamic volume of the black hole, $\mathcal{A}$ is its area, and $\Omega_{d-2}$ is   the dimensionless volume of a
`unit $(d-2)$-ball' in the $d$-dimensional spacetime.

The rest of our paper is organized as follows. In the next section we review the charged C-metric, discuss its thermodynamic behaviour,  and
look closer at the admissible parameter space of the  solution, which will provide key insights as to why the swallowtail snaps. Section~\ref{sec: phase transitions} is devoted to the study of the  phase behaviour.  We end with summary and conclusions in section~\ref{sec: conclusion}.

\section{Charged AdS C-metric and its thermodynamics}
\label{sec: charged c-metric}

\subsection{Solution}
The accelerating charged AdS C-metric can be written as follows \cite{Podolsky:2002nk,Hong:2004dm,Griffiths:2005qp}
\ba\label{eq: metric}
ds^2 &=& \frac{1}{\Omega^2}\bigg[ -\frac{fdt^2}{\alpha^2} + \frac{dr^2}{f} + r^2\bigl(\frac{d\theta^2}{g} + {g\sin^2\theta} \frac{d\phi^2}{K^2}\bigr)\bigg]\,, \nonumber\\
F&=&dB,\qquad B=-\frac{e}{\alpha}\Big(\frac{1}{r}-\frac{1}{r_+}\Big) dt\,,
\ea
where we have chosen a gauge in which the electrostatic potential vanishes on the black hole horizon, and where
\ba
f&=&(1-A^2r^2)\Bigl(1-\frac{2m}{r}+\frac{e^2}{r^2}\Bigr)+\frac{r^2}{l^2}\,,\nonumber\\
g&=&1+2mA\cos\theta+e^2A^2\cos^2\!\theta\,,
\ea
with $A$ being the acceleration parameter, $\Lambda = -3/l^2$ the cosmological constant, $m$ the mass parameter and $e$ the charge parameter.  The time coordinate has been rescaled by the  parameter $\alpha$,
\be\label{alp-eq}
\alpha=\sqrt{\Xi(1-A^2 \ell^2\Xi)}\,, \qquad  \Xi=1+e^2A^2\,,
\ee
 in order to ensure a consistent variational principle  and a correct normalization of the timelike Killing vector at infinity \cite{Anabalon:2018ydc,Anabalon:2018qfv}. 
 The conformal factor
\beq \Omega=1+Ar\cos\theta
\eeq
determines the boundary of the AdS spacetime, and
the parameter $K$ encodes information about the conical deficit on the south and north poles, so that $\phi \in [0,2\pi]$.

  The metric and gauge potential \eqref{eq: metric} satisfy the Einstein-Maxwell equations everywhere except along the polar axes $\theta=\theta_+ = 0$ and $\theta=\theta_- = \pi$, where there must exist a string of stress-energy in order to offset the conical deficit about these axes.
  This can be done by introducing cosmic strings whose tensions on the polar axes are
\beq
\mu_{\pm}=\frac{1}{4}\Bigl(1-\frac{\Xi\pm2mA}{K}\Bigr)\,,
\eeq
 and are related to the
 conical deficits, $\delta_\pm$ by $\delta_\pm=8\pi \mu_\pm$.  Thus we have the following range for the tensions:
\be
\mu_\pm\in[0,{1}/{4})\,,
\ee
with the upper limit corresponding to a conical deficit of $2\pi$.
Defining  further
\beq
\label{eq: K}
K_{\pm}=g(\theta_{\pm})=\Xi\pm 2mA\,,
\eeq
we can by an appropriate choice of $K=K_+$ or $K=K_-$ respectively set either of $\mu_+$ or $\mu_-$ to zero, but not both.

If the black hole has sufficiently slow acceleration (see discussion below) there will be a single (black hole) horizon.  This constrains the parameter space and allows for a single temperature and a consistent thermodynamics. We turn now to consider this situation.

\subsection{Thermodynamics}

Before considering the conditions required for the slow acceleration, let us present the
thermodynamic quantities associated with the charged accelerating black hole \cite{Anabalon:2018qfv}
\ba\label{AdSthermo}
M&=& \frac{m(1-A^2 \ell^2\Xi)}{K\alpha}  =\frac{m\sqrt{1-A^2 \ell^2\Xi}}{K\sqrt{\Xi}}\,, \label{mass}\\
T&=& \frac{f'_+}{4\pi\alpha}\,, \quad
S=\frac{\pi r_+^2}{K(1-A^2r_+^2)}\,, \label{TS}\\
Q&=& \frac{e}{K}\,,\;\; \Phi=\frac{e}{r_+\alpha}\,,\quad
P = \frac{3}{8\pi \ell^2}\,,  \label{QPhiP}\\
V &=& \frac{4\pi}{3K\alpha} \left [ \frac{r_+^3}{(1-A^2 r_+^2)^{2}}
+ mA^2\ell^4\Xi\right]\, , \label{vol}\\
\lambda_\pm &=& \frac{r_+}{\alpha(1\pm Ar_+)} - \frac{m}{\alpha\Xi}
\mp \frac{A \ell^2 \Xi}{\alpha}\,.  \label{tension}
\ea
They satisfy
\ba\label{flaw}
\delta M&=&T\delta S+\Phi \delta Q+V\delta P -\lambda_+\delta \mu_+
-\lambda_-\delta \mu_- \,,\nonumber\\
\label{Smarr}
M&=&2(TS-PV)+\Phi Q\,,
\ea
which are respectively the first law and   Smarr relation.  The necessity for introducing the last two work terms to the first law was first demonstrated in \cite{Appels:2017xoe}.  
Examples of where string tensions do vary  include ``capture of cosmic string by a black hole'' and an ``axisymmetric merger'' of two accelerating black holes, each carrying its own cosmic string.

In what follows we explicitly make the choice $K=K_+$, so that
\be
\label{eq: mu}
\mu_+=0\,,\quad \mu=\mu_-=\frac{mA}{K_+}\,.
\ee
In other words, only one string (located at the south pole) pulls on the black hole, which is completely regular on the north pole.
 The first law then takes the form \eqref{two}.  In the sequel  we will hold the remaining tension fixed, and probe the effect of its particular value on the thermodynamic phase behavior.

\subsection{Parameter space}
\label{sec: parameter space}

\begin{figure}[t]
\centering
\subfigure[]{\includegraphics[scale=0.55]{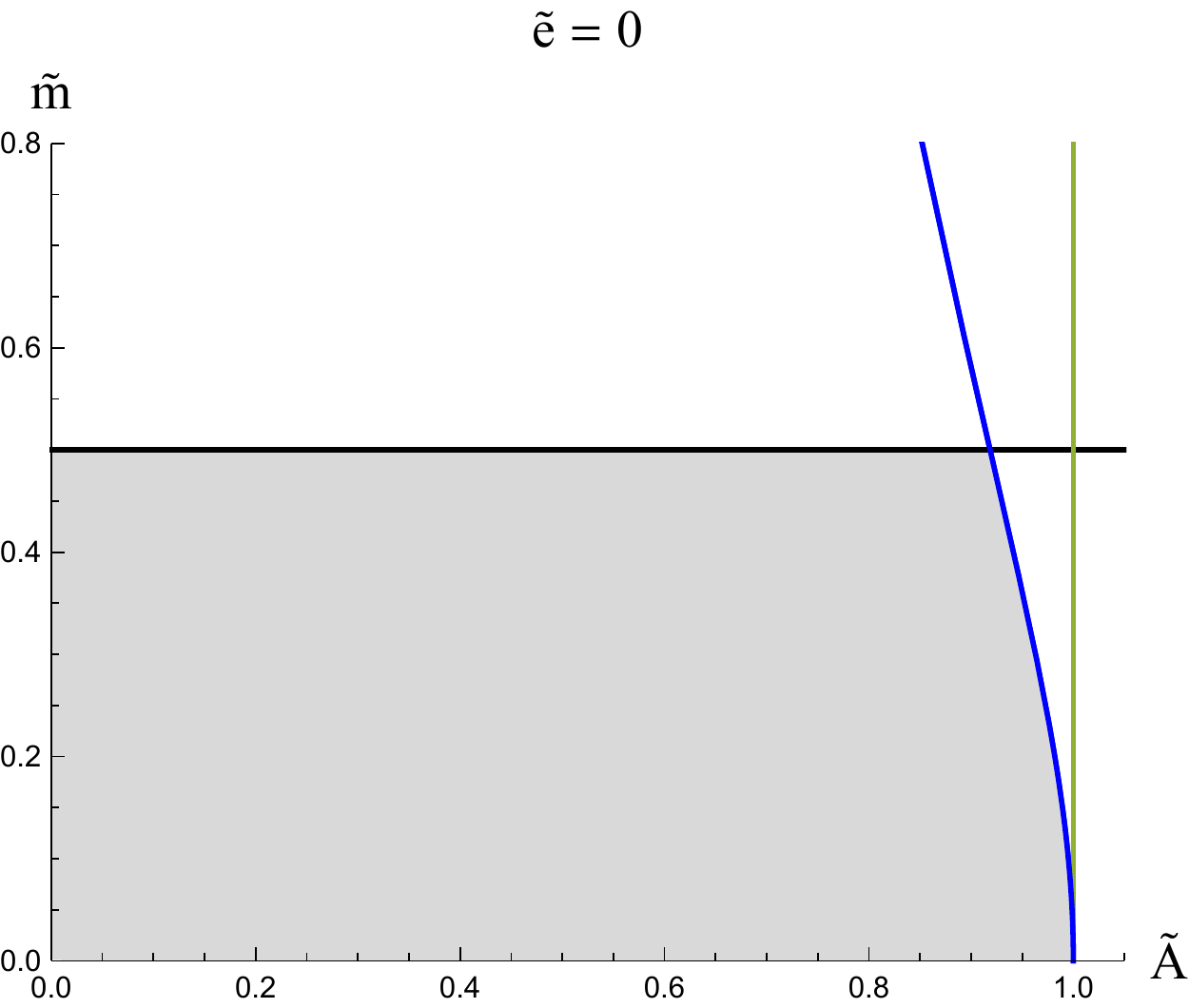}} \quad
\subfigure[]{\includegraphics[scale=0.55]{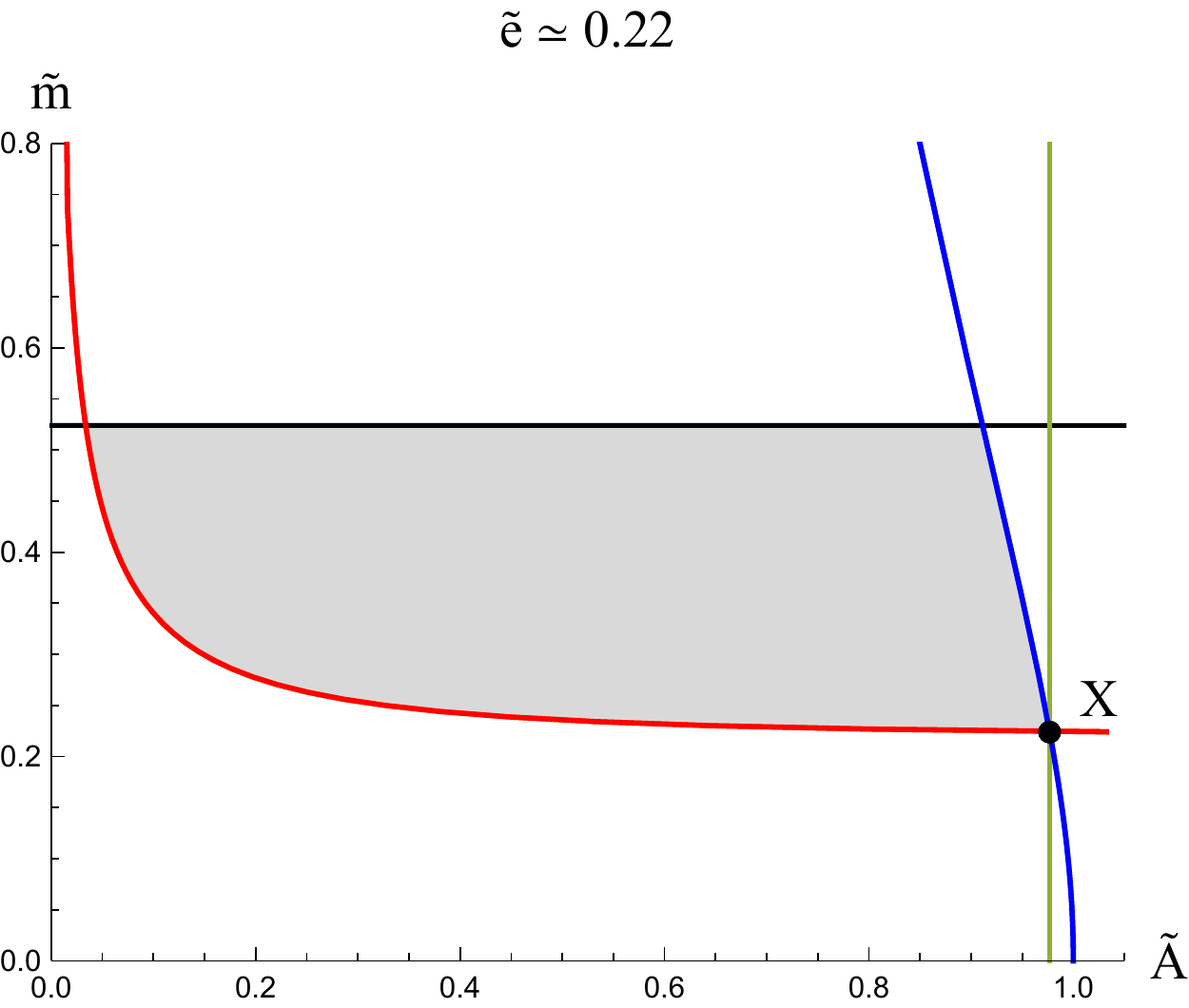}}
\caption{ \textbf{Parameter space.} The admissible parameter space (denoted by shaded areas) is displayed in the $\tilde m$ vs. $\tilde A$ plane for fixed  $\tilde e = 0$ (left) and  $\tilde e \approx 0.22$ (right). 
Horizontal black curves outline the boundary of positive $g$. The {red} curve is the boundary for the existence of black holes in the bulk, with extremal black holes sitting on the curve. The blue curve forming the right boundary of the admissible region gives the no-acceleration horizon condition, and the green vertical line corresponds to $\alpha=0$.  A non-trivial point $X$ is observed for $0<\tilde e\leq\tilde e_M$. For $\tilde e =0$ the analogous point  is situated at the
at $\tilde A=1, \tilde m=0$ corner of the left diagram, which  corresponds to  pure AdS spacetime.}
\label{fig: parameter space}
\end{figure}

Before discussing the possible phase transitions, we must determine the admissible space of the parameters appearing in the metric~\eqref{eq: metric} that ensure our problem is well-posed. Clearly, we have a number of conditions to impose: i) positivity of the function $g$ over the range $\theta\in [0,\pi]$, ii) existence of a black hole in the bulk, and  iii) validity of the derived thermodynamics. The latter entails working in the slow acceleration regime~\eqref{slow} and~\eqref{Att}. This ensures that $\alpha$ is  positive  and that there is a single horizon whose temperature is given by its surface gravity.

In order to discuss these conditions, we introduce the new coordinates
\begin{equation}
x=\frac{1}{Ar}\,,\quad y=\cos\theta
\end{equation}
so that the conformal boundary is situated at $x=-y$. We also consider the dimensionless quantities
\begin{equation}
\label{eq: dimless}
\tilde m=mA\,,\quad \tilde e=eA\,,\quad \tilde A=Al\,, \quad  \tilde r=r/l\,.
\end{equation}
 Using these variables, and the definition of $\mu$, $K$, and $Q$, we find the following relations:
 \begin{equation}\label{useful}
\tilde A=\frac{\tilde e(1-2\mu)}{(1+\tilde e^2)Q/l}\,,\quad \tilde m=\frac{\mu(1+\tilde e^2)}{1-2\mu}\,.
\end{equation}

{\em Signature of $g$.}
In order to have the right metric signature $(-,+,+,+)$, we require $g>0$ for $y\in [-1,1]$. This yields
\be\label{mtt}
{\tilde m < }\begin{cases}
\frac{1}{2}(1+\tilde e^2)\quad \mbox{for}\quad |\tilde e|<1\,,\\
|\tilde e|\quad \mbox{for}\quad | \tilde e|>1\,,
\end{cases}
\ee
the boundaries of which give the  black horizontal lines in Fig.~\ref{fig: parameter space}.

 {\em Existence of a bulk black hole.}
Demanding that the spacetime admits a black hole and not a naked singularity, we require that a horizon exists in the bulk, that is $f$ has  at least one root  {$r_+$
 in the range $r\in (0, 1/A)$ preserving the signature for  $r_+ < r < 1/A$.} 
 To find this we can appeal to the condition for an extremal black hole
\be
 {f(\tilde r_e)}=0=f'(\tilde r_e)\,,
\ee
 or in other words, that $f$ has a double root {$r_+ = r_e$, corresponding to the coincidence of the inner and outer black hole horizons.}
Solving these two equations yields
 $\tilde A=\tilde A(\tilde r_e, \tilde e)$ and $\tilde m=\tilde m(\tilde r_e, \tilde e)$, which can be plotted parametrically in terms
 of $\tilde r_e$  for a fixed $\tilde e$. The resultant curve provides a lower bound for the existence of the black hole,
 and is displayed in Fig.~\ref{fig: parameter space}b  by the {red} line, denoting the extremal limit. Above this line, a black hole (with two horizons) is present, whereas no black hole exists below it.

{\em Validity of thermodynamics.}
To ensure that  the thermodynamic quantities \eqref{AdSthermo} are well defined,
 it is necessary to have
 $\alpha>0$, which in turn imposes
\beq\label{slow}
1-A^2\ell^2\Xi>0\,,
\eeq
or
\be
\tilde A<\frac{1}{\sqrt{1+\tilde e^2}}\,.
\ee
The boundary is displayed in Fig.~\ref{fig: parameter space} by the vertical green lines.

On the other hand, the sufficient condition for the slow acceleration regime is that $f$  does not develop any roots on the boundary (neither acceleration nor cosmological  horizons are present  \cite{Podolsky:2003gm}). Since the
conformal  boundary is situated at $x=-y$, the metric function $f$ develops a root on the boundary when
\be\label{noboundaryhorizon}
f(x=-y)=0=f'(x=-y)\,,
\ee
for some $y\in [-1,1]$. This
yields the following relations:
\begin{equation}
\label{Att}
\tilde m=\frac{y(1+2\tilde e^2 y^2-\tilde e^2)}{1-3y^2}\,,\quad \tilde A=\pm \frac{\sqrt{(1-\tilde e^2 y^2)(1-3y^2)}}{(1-y^2)(1-\tilde e^2 y^2)}\,.
\end{equation}
The corresponding parameter space can be plotted parametrically, for $y\in [-1,1]$;  it corresponds to the blue curves forming the right-hand boundaries of the admissible regions in Fig.~\ref{fig: parameter space}.

 Note that the condition $\alpha>0$, for which the thermodynamic quantities \eqref{AdSthermo} are well defined functions, is weaker than the requirement of the slow acceleration regime---the blue curve cuts away an additional piece of the admissible region of the parameter space.  One might suspect that the thermodynamic quantities in this removed  region would correspond to the characteristics of a rapidly accelerating black hole. Even if so, the phase transition interpretation in this regime would be questionable as there are additional horizons present in the spacetime.  We note, however,  that a proposal  \cite{Kubiznak:2015bya} to treat  additional  horizons (in that case de Sitter) as `independent thermodynamic systems' that do not apriori affect the phase transitions due to the black hole horizon has been recently put forward.

 {\em Point $X$ and a summary of constraints.}
The above constraints   dictate the admissible parameter space in the dimensionless $(\tilde A,\tilde m,\tilde e)$ plane. Two dimensional slices of this three-dimensional parameter space  can easily be displayed;  see Fig.~\ref{fig: parameter space} for two  examples of  $\tilde e$: $\tilde e=0$ (left) and $\tilde e\approx 0.22$ (right). The full admissible parameter region is a union of such slices.

 For a non-trivial $\tilde e$ we note the presence of a `{\em point $X$}', the only point where the $\alpha=0$ line forms the boundary of the admissible region. As we shall see this point plays a crucial role for the existence of snapping swallowtails. Point $X$  is characterized by the intersection of the following three curves: the $\alpha=0$ curve, the extremal black hole curve, and the slow acceleration curve. Since the intersection of any two is sufficient for finding $X$, this point is (for example) given by
\be\label{XXX}
\alpha=0\,,\quad f(\tilde r_e)=0=f'(\tilde r_e)\,,
\ee
which yields
\ba\label{Xpoint}
\tilde A&=&\frac{1}{\sqrt{1+\tilde e^2}}\,,\quad \tilde m=|\tilde e|\sqrt{1+\tilde e^2}\,,\nonumber\\
\tilde r_e&=&\frac{\sqrt{1+6\tilde e^2+5 \tilde e^4}-1-\tilde e^2}{2|\tilde e|}\,.
\ea
The definition of $Q$ and $\mu$ then gives
\be\label{Xpoint2}
Q=\frac{\tilde e \sqrt{1+\tilde e^2}}{1+\tilde e^2+2|\tilde e|\sqrt{1+\tilde e^2}}\,,\quad \mu=\frac{|Q|}{l}\,,
\ee
with the latter equivalent to the critical pressure, \eqref{Pt1},
\be\label{Pt2}
P_t=\frac{3\mu^2}{8\pi Q^2}\,.
\ee

As $\tilde e$ increases the point X `travels upwards' and the admissible parameter region shrinks, until at a maximum   value $\tilde e_M$ the whole parameter space shrinks to one point.  This occurs when  all boundary curves intersect, that is for
\be
 |\tilde e_M|=\frac{\sqrt{3}}{3} \qquad  \tilde m_M= \frac{2}{3} \qquad  \tilde A = \frac{\sqrt{3}}{2} \, .
\ee
For larger $\tilde e$  there is no physically admissible region.    Hence $\tilde e$ is constrained to the range $\tilde e\in [0, \tilde e_M]$, which in turn implies that the mass $\tilde m\in[0,\tilde m_M]$.
We note that these ranges imply that only the upper formula in \eqref{mtt} and the plus sign in \eqref{Att} are applicable.

\subsection{Mini-entropic black holes}

{The point $X$, characterized above by \eqref{XXX} and \eqref{Xpoint}--\eqref{Pt2}, would correspond, via
\eqref{mass}--\eqref{tension}, to a black hole of vanishing mass but finite radius and entropy, as well as infinite volume, and  potential\footnote{ Note that the formula \eqref{Xpoint} admits a smooth limit $\tilde e\to 0$, in which case we recover $\tilde A=1, \tilde m=0=\tilde r_e$, corresponding to   empty AdS space.}.
Of course, this object cannot actually be physically realized since $X$ lies outside of the admissible region in parameter space (and thence the above stated thermodynamics does not apply).  However it is possible to come arbitrarily close to it, for example following the slices of constant charge, pressure, and string tension, as displayed in Fig.~\ref{fig: slicing} above.

\begin{figure}
    \centering
    \includegraphics[scale=0.65]{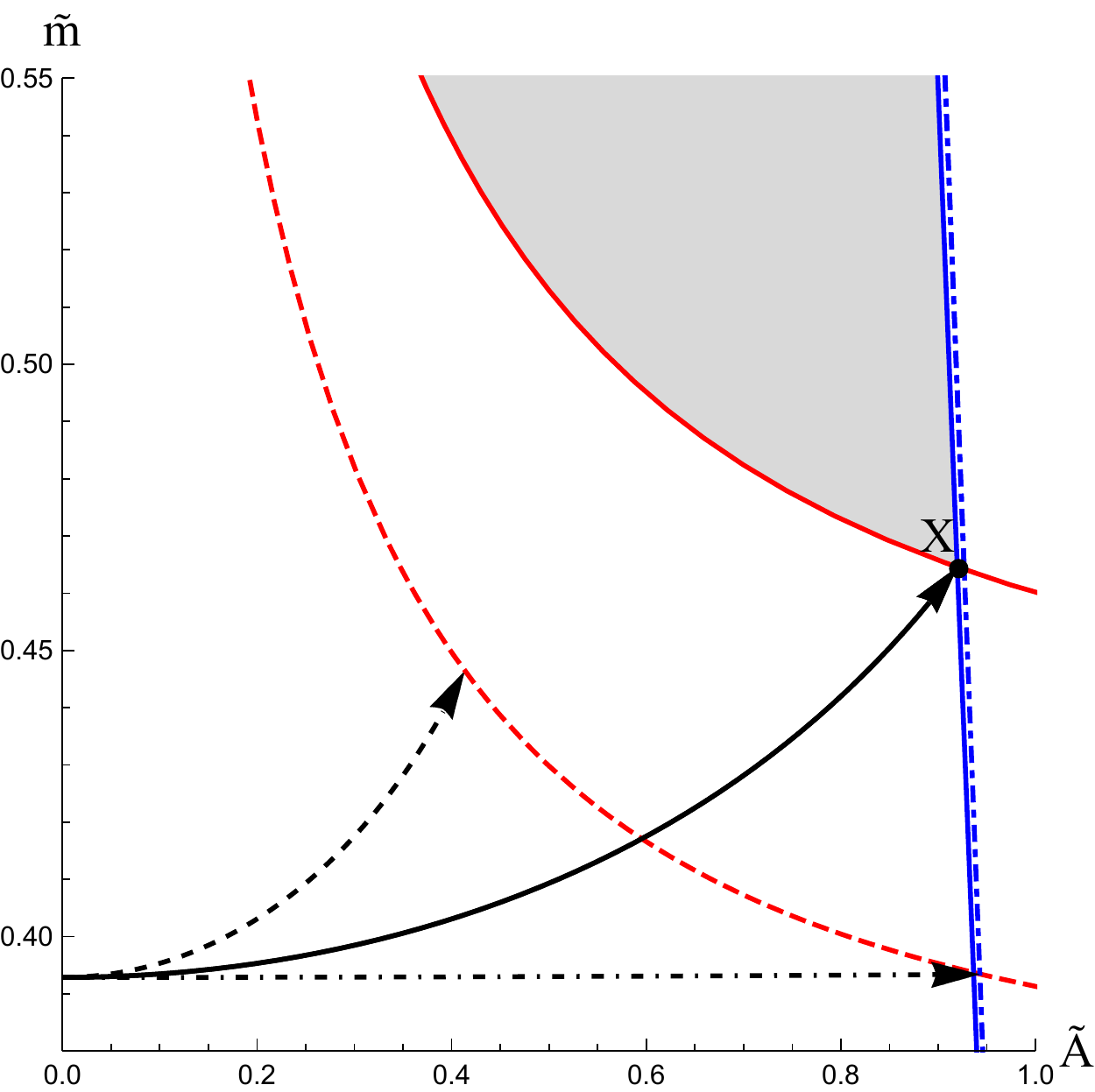}
    \caption{\textbf{Swallowtail slicing of the parameter space.} We display in black a projection of three different free energy ``swallowtail''
    curves in the $(\tilde m,\tilde A)$ parameter space for  $\mu = 0.22$ and $Q=1$, with the associated critical pressure $P_t$ and the associated point $X$;        the shaded region is the admissible region for the slice 
    $\tilde e\approx 0.4272$,
  the plane on which $X$ lies.
  The solid black curve corresponds to the critical swallowtail with $P=P_t$, it terminates at $X$.
     The dashed curve corresponds to $P>P_t$: any such curve terminates on an extremal black hole curve denoted (similar to Fig.~\ref{fig: parameter space}) by the red dashed curve of some $\tilde e$ slice (for small enough pressures this is a true swallowtail).   The dot-dash curve characterizes $P<P_t$ and terminates on a slow black hole curve denoted by dot-dash blue curve; such curves correspond to `Schwarzschild-AdS'-like behavior of the snapped free energy.}
    \label{fig: slicing}
\end{figure}

Computing the isoperimetric ratio $\mathcal{R}$ in \eqref{eq:ipe-ratio}, we see that it diverges
as the point $X$ is approached, where we have taken $\Omega_2 = 4\pi/K$ as the dimensionless volume of a
`unit ball', as determined by the $r_+$-independent metric conformal to the metric of a constant
$(t,r)$ hypersurface of \eqref{eq: metric}.   From the perspective of the reverse isoperimetric inequality $\mathcal{R} \geq 1$ \cite{Cvetic:2010jb}, black holes in the vicinity of $X$ are  \textit{mini-entropic}:  their volume diverges and their area is finite.  This is in contrast to  super-entropic black holes \cite{Hennigar:2014cfa},  whose entropy exceeds the maximum implied by the black hole volume ($\mathcal{R} < 1$).

\section{Phase transitions}
\label{sec: phase transitions}

As per usual, to uncover the thermodynamic behaviour of the system, we study the free energy, which is now also a function of the string tension $\mu$,
\be
F=M-TS=F(T,P, Q, \mu)\,.
\ee
Thermodynamic equilibrium corresponds to the global minimum of   $F$.
Non-analytic behaviour of this minimum indicates the presence of phase transitions.

\subsection{ Snapping swallowtails}

\begin{figure}
    \centering
    \subfigure[]{\includegraphics[scale = 0.75]{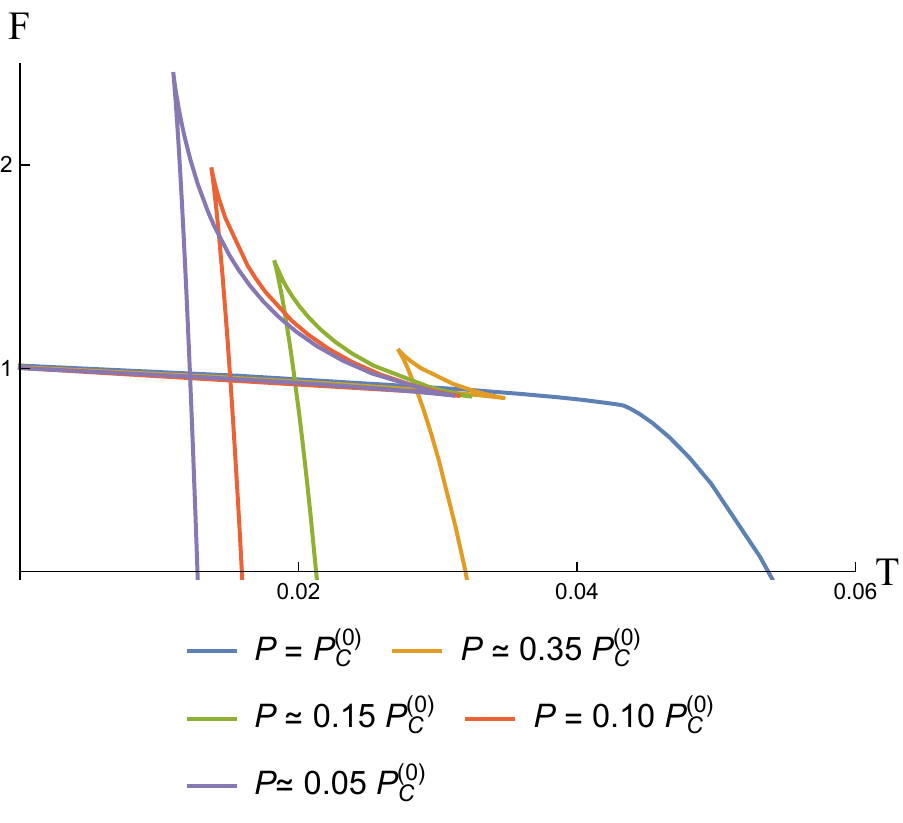}}\quad
    \subfigure[]{\includegraphics[scale = 0.80]{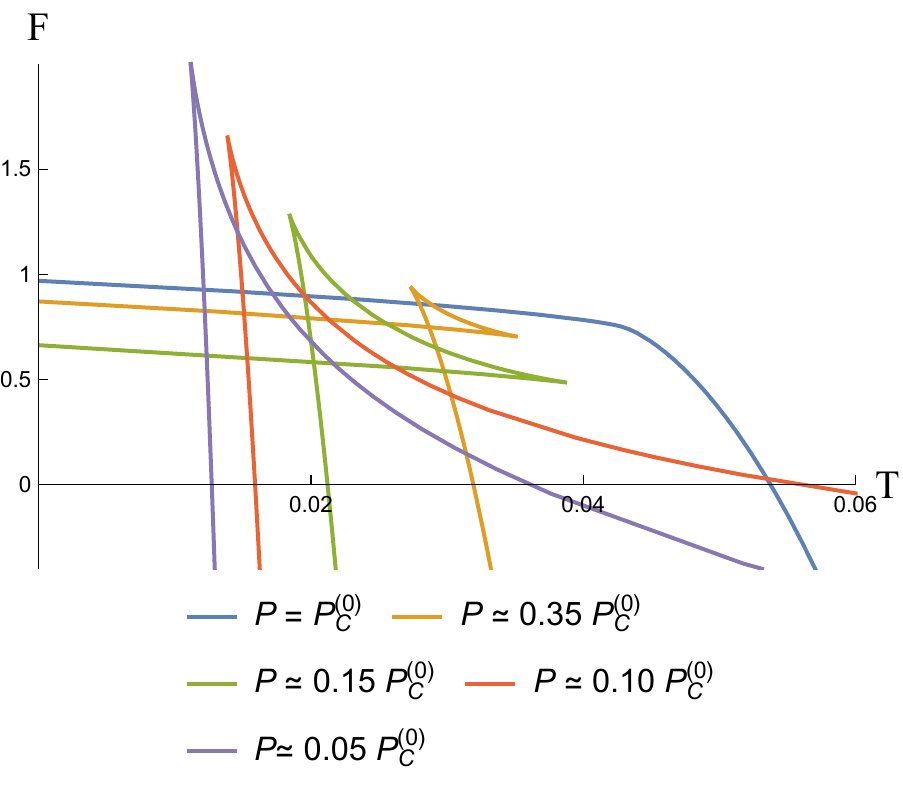}}
    \caption{   \textbf{Free energy.} {\em Left:} {A plot of free energy as a function of temperature
    for non-accelerating ($\mu=0$) charged black holes for various values of the pressure, measured in terms of the critical pressure $P_c^{(0)}$.
    {\em Right:} A similar plot, but now for accelerating charged black holes with $\mu = 0.05$; the pressures are the same as in the left figure.} Whereas on the left figure the swallowtails exist for arbitrary small pressures, on the right the swallowtail disappears for $P<P_t$, see e.g. red curve where the swallowtail already snapped.
    In both pictures we have set $Q=1$.
    }
    \label{fig: phase transition}
\end{figure}

As discussed in the introduction,
charged AdS black holes without acceleration exhibit   Van der Waals-like phase behaviour \cite{Kubiznak:2012wp}:  for pressures  $P<P_c$ one observes   swallowtail behaviour of the free energy---corresponding to the first order small black hole/large black hole phase transition, as illustrated in the left part of  Fig.~\ref{fig: phase transition}. The curve is smooth for $P>P_c$, develops a kink at $P=P_c$, and then a swallowtail for $P < P_c$
which grows in size as pressure further decreases.
For any given swallowtail, beginning at the left at $T=0$, for increasing temperature the black hole
has increasing radius.  Following the curve rightward to its first cusp, then leftward and up to its second cusp,
and then rightward again down the steeply negative slope, the radius of the black hole monotonically increases.
However the global minimum of the free energy experiences a discontinuity at the intersection, at which
it is thermodynamically favourable for the small black hole to undergo a first-order phase transition to a large black hole.

In the presence of acceleration the situation is much more interesting. As depicted in
the right-hand diagram of  Fig.~\ref{fig: phase transition},  for small  tensions and moderate pressures the swallowtail behaviour is preserved.   However, contrary to the non-accelerating case, the swallowtail ceases to exist for $P<P_t$. Instead,  as $P$ decreases through $P_t$, the swallowtail `snaps':  the small black hole branch disappears together with the extremal black hole, and  re-appears as a new branch of unstable high temperature black holes (that are mini-entropic for sufficiently high temperatures). The  resultant free energy diagram for $P<P_t$ is reminiscent of that of the `Schwarzschild-AdS' black hole (see Fig.~\ref{fig: snap} for more details). We stress, however, that in this case no Hawking--Page transition can exist, as there is no radiation phase with non-trivial charge $Q$ and non-trivial string tension $\mu$.

\begin{figure}
    \centering
    \includegraphics[scale=0.65]{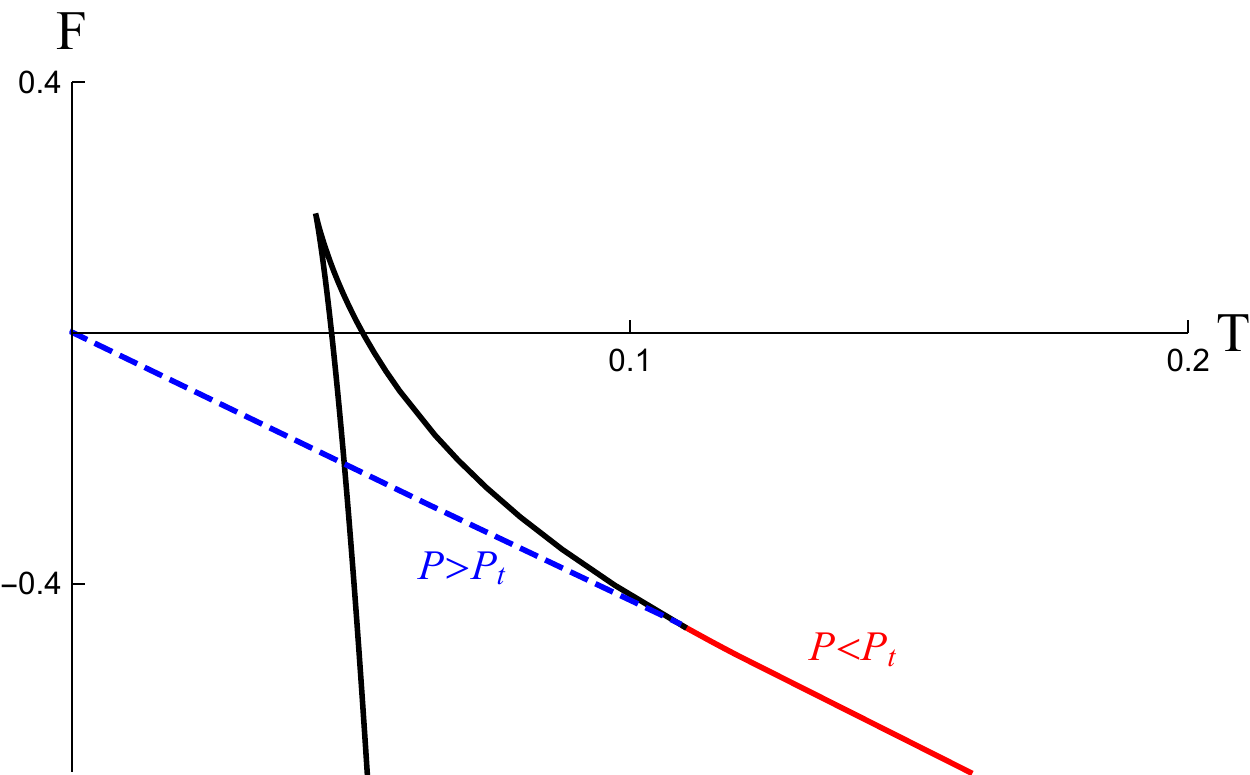}
    \caption{\textbf{Snapping swallowtail at $P\approx P_t$.} We display the qualitative change of the behaviour of the free energy around $P\approx P_t$ for $\mu=0.1$ and $Q=1$. The blue dashed curve together with the solid black curve show the behaviour of $F$ for $P\gtrsim P_t$ where the swallowtail is present. At $P_t$ the swallowtail snaps, the blue dashed curve disappears and a new branch of black holes displayed by the red curve appears. The black holes on the dashed blue curve map to the black holes on the solid red curve in such a way that the cusp point remains invariant  and the extremal black hole located on the blue curve at $T=0$ maps to the (red curve) mini-entropic black hole at $T=\infty$.  At $P\lesssim P_t$ we observe the free energy,  reminiscent of that of the Schwarzschild-AdS black hole, given by a union of red and black curves. }
    \label{fig: snap}
\end{figure}

 To get an intuitive feeling as to why the swallowtail snaps, let us study the $P>P_t$, $P=P_t$, and $P<P_t$ `swallowtails' of the free energy (displayed in Fig.~\ref{fig: phase transition}) from a perspective of the corresponding parameter space slicing. 
Such curves are characterized by fixed $\mu$ and $Q$, and of course fixed pressure, and
 give rise to  corresponding curves in the parameter space $(\tilde m,\tilde A, \tilde e)$. In Fig.~\ref{fig: slicing} we display
the orthogonal projection of these curves onto a fixed $\tilde e$ plane.
On one end such curves asymptote to $\tilde A=0$ (which would happen only for $\tilde e=0$) corresponding to the stable large black holes, on the other end they terminate either on an extremal black hole (red dashed) curve, at the point $X$, or on the slow acceleration blue curve; in general each happening for a different value of $\tilde e$.

When $P=P_t$ (displayed in Fig.~\ref{fig: slicing} by a solid black curve) the swallowtail terminates at the point $X$ and represents the critical slice\footnote{That this slice seems to lie beyond the admissible region is simply an artifact of displaying the 2d projections of the full 3d parameter space. Since these regions get bigger as $\tilde e$ decreases, all points on the slice actually belong to the admissible region. }.
For $P_c > P > P_t$  the (dashed black) curve lies to the left of the critical slice and represents a true swallowtail, terminating on an extremal black hole curve at some value of $\tilde e$. On the other hand, if $P<P_t$ the (dot-dash black) curve lies to the right of the critical slice and necessarily terminates on the slow acceleration boundary curve for some $\tilde e$. Of course, this corresponds to the already snapped swallowtail.

Summarizing,  it is the `crossing' of the point $X$ in parameter space, together with the existence of the slow acceleration boundary, that is responsible for  the swallowtail snapping.

\subsection{ Zeroth order phase transition \& bicritical point}

 \begin{figure}
    \centering
    \includegraphics[scale=0.65]{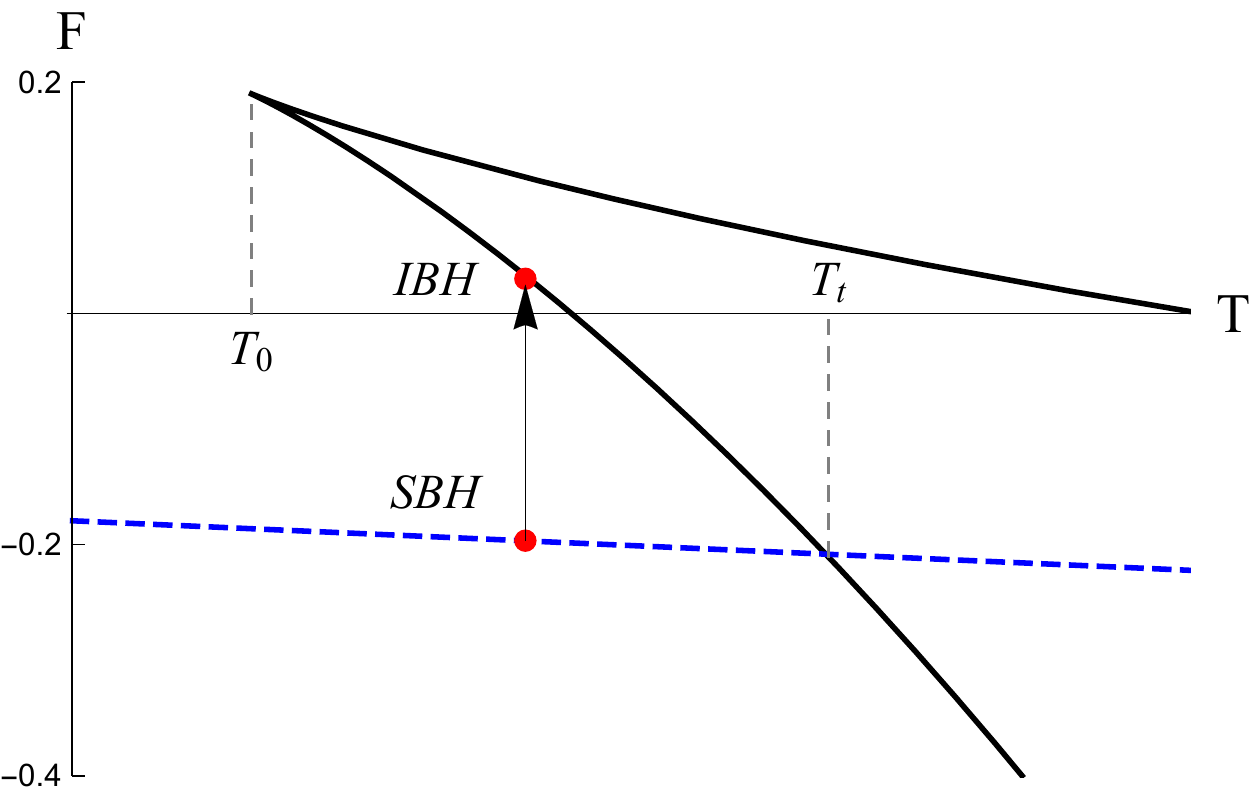}
    \caption{\textbf{Zeroth order phase transition.} Due to the swallowtail snap at $P=P_t$, the entire branch of stable low temperature black holes (blue dashed curve) disappears. This gives rise to a no black hole region for temperatures $T<T_0$. At the same time the global minimum of the free energy in between $T_0$ and $T_t$, `jumps upwards' from the small black hole branch to the intermediate black hole branch, as displayed in the figure for a single black hole of a given temperature. }
    \label{fig: zeroth}
\end{figure}

The snap of the swallowtail at $P=P_t$ has the following rather interesting consequence illustrated in Fig.~\ref{fig: zeroth}. As we lower the pressure through $P_t$, the entire branch of stable small low temperature black holes present for $P\gtrsim P_t$ disappears.
Consequently for $P\lesssim P_t$ there are no longer any black holes below $T_0$, with $T_0$ being the temperature of the upper cusp of the critical swallowtail. At the same time the global minimum of the
\begin{figure}[htp!]
     \centering
        \subfigure[$\,\mu = 0.05$]{\includegraphics[scale=0.65]{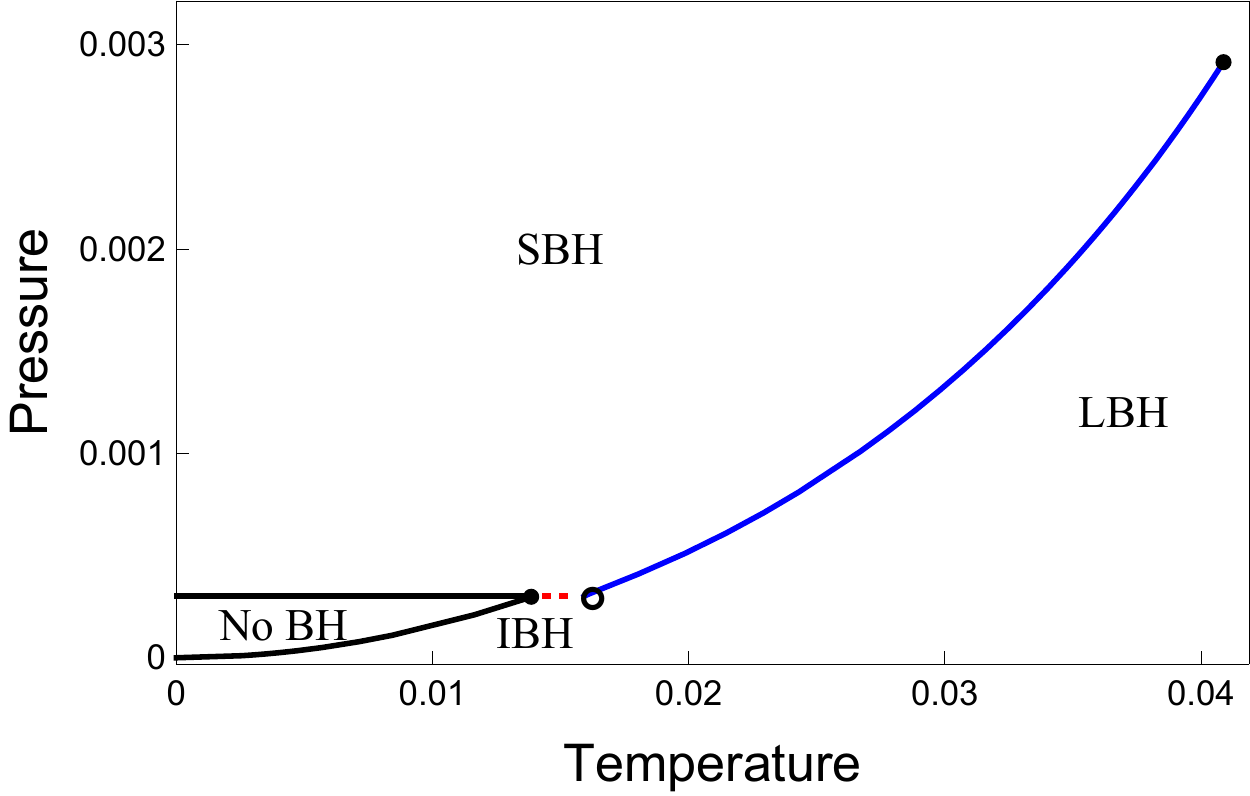}}
        \subfigure[$\,\mu = 0.15$]{\includegraphics[scale=0.65]{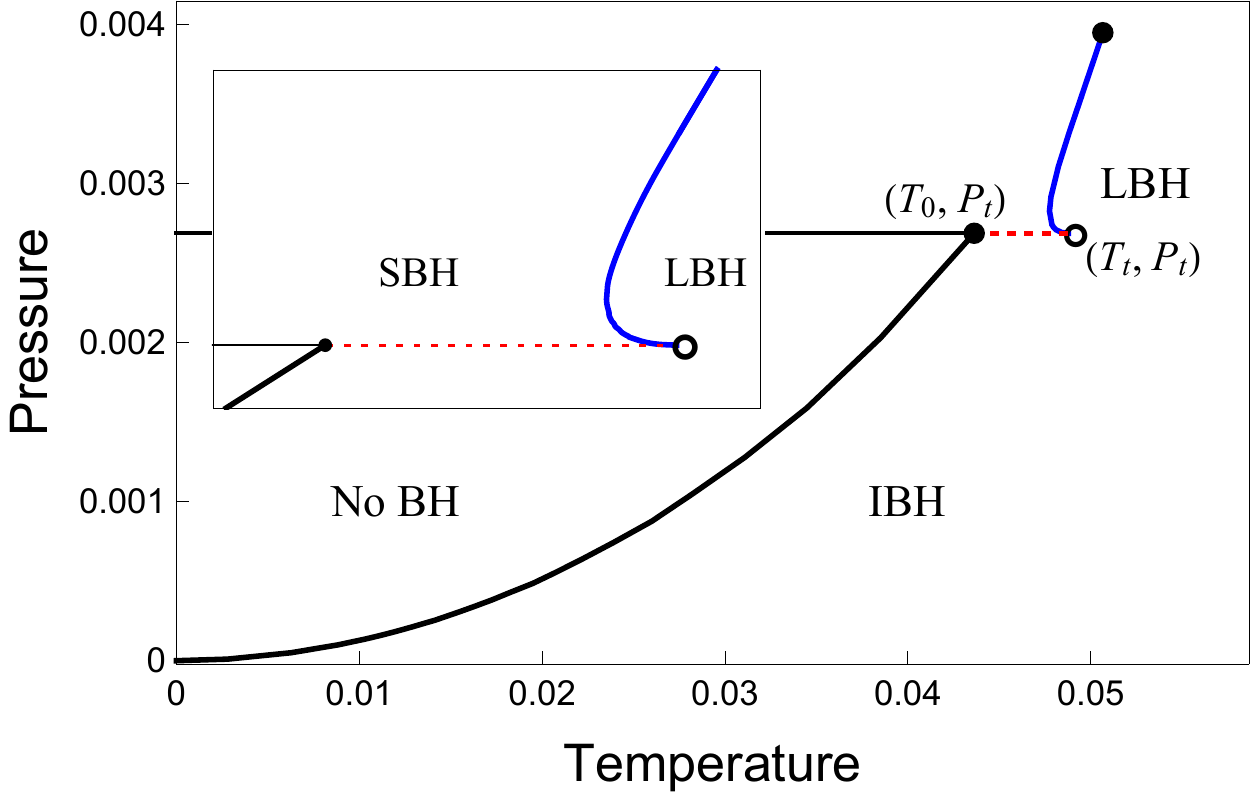}}
        \subfigure[$\,\mu \approx 0.25$]{\includegraphics[scale=0.65]{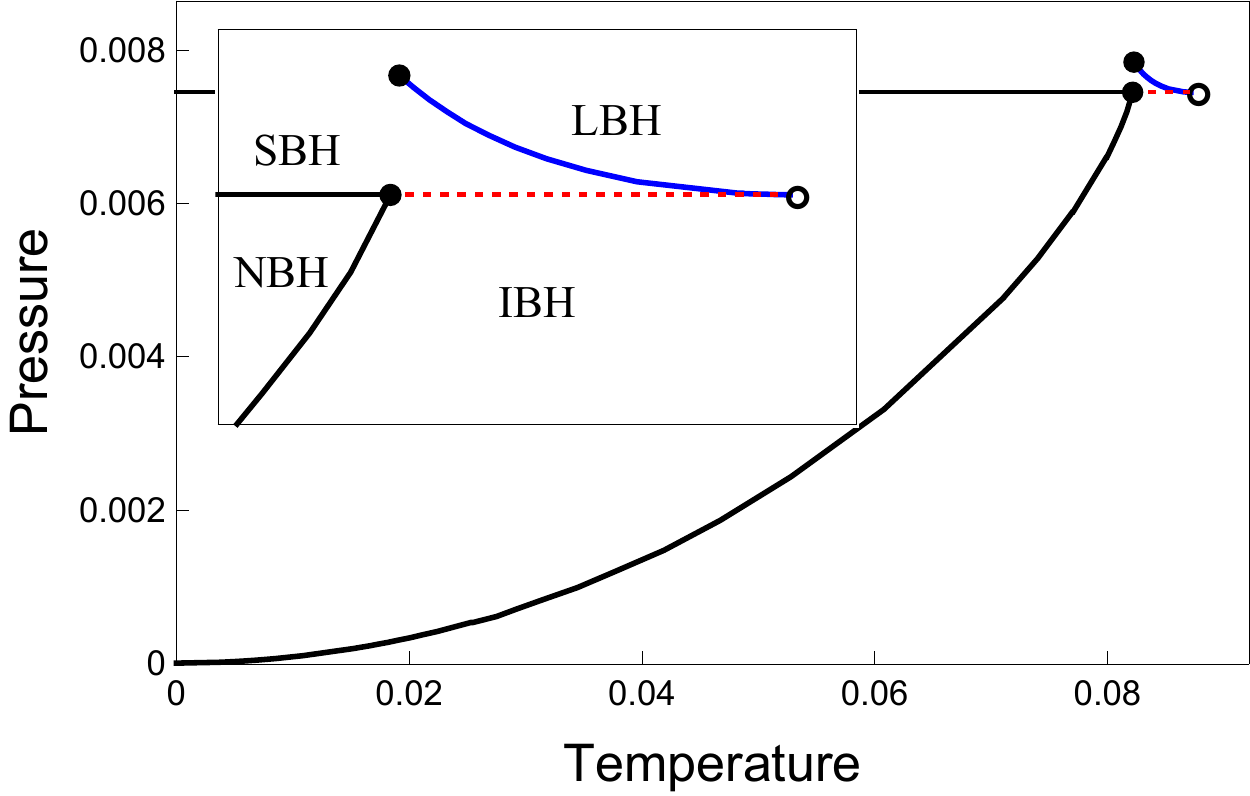}}
        \caption{ {\bf Phase diagrams.}
        $P-T$ diagrams are displayed for various string tensions. The blue curve indicates the coexistence line of the first order phase transition. It terminates at a critical point (denoted by a black solid circle) on one end and at a bicritical point (characterized by $(T_t,P_t)$ and denoted by empty circle) on the other end. A zeroth-order phase transition coexistence curve  emanates from the bicritical point,  indicated by the dotted red line. The black curves bound a region for
        which  no slowly accelerating black holes exist, the `no black hole' region. The inset of the second diagram clearly illustrates the presence of a reentrant phase transition with pressure as the control parameter. Note also the slope of the blue coexistence line in the third diagram, which is `opposite to' what happens for the non-accelerated case.}
         \label{fig: RPT}
\end{figure}
free energy in between $T_0$ and $T_t$, `jumps upwards' from the small black hole branch to the intermediate black hole branch, as schematically illustrated for a single black hole in Fig.~\ref{fig: zeroth}. Here $T_t$ is the temperature of the bottom intersection of the critical swallowtail.

In other words, in between $T_0$ and $T_t$ we observe a \textit{zeroth-order phase transition}, as the global minimum of the free energy experiences  a finite jump across $P_t$. Increasing the temperature from $T_0$, this jump gets smaller and smaller and finally disappears at $T=T_t$. Thus, we have a special point, at
\be
(T_t,P_t)\,,
\ee
where the coexistence line of the zeroth order phase transition characterized by $P_t$ and $T\in (T_0, T_t)$ terminates and joins the first order coexistence line of the small black hole/large black hole phase transition. We call this point the bicritical point.

\subsection{Phase diagrams}

 Having understood the free energy, we can now plot the phase diagrams, displayed for various tensions in Fig.~\ref{fig: RPT}. Consider first a small string tension $\mu=0.05$, Fig.~\ref{fig: RPT}a. Similar to the non-accelerating case the diagram features the first order phase transition coexistence line (displayed by the blue curve) and the critical point (solid circle) where the first order coexistence line terminates and the transition becomes  second order. A novel feature, when compared to Fig.~\ref{fig1}, is the existence of a bicritical point (empty circle) and the associated zeroth order phase transition (red dashed curve). We also observe the existence of a no black hole region caused by the fact that low temperature  slowly accelerating black holes no longer exist below $P_t$. \
 Note that the situation is physically very different from the Hawking--Page transition \cite{Hawking:1982dh}, where the `no black hole region' is replaced by a radiation phase with lower free energy. In our ensemble of fixed charge and fixed tension of the semi-infinite string, no such radiation phase exists. It remains to be seen whether some novel phase of solutions (preserving the ensemble conditions) may exist in this region and be thermodynamically preferred.

\begin{figure}[htp]
    \centering
    \includegraphics[scale=0.65]{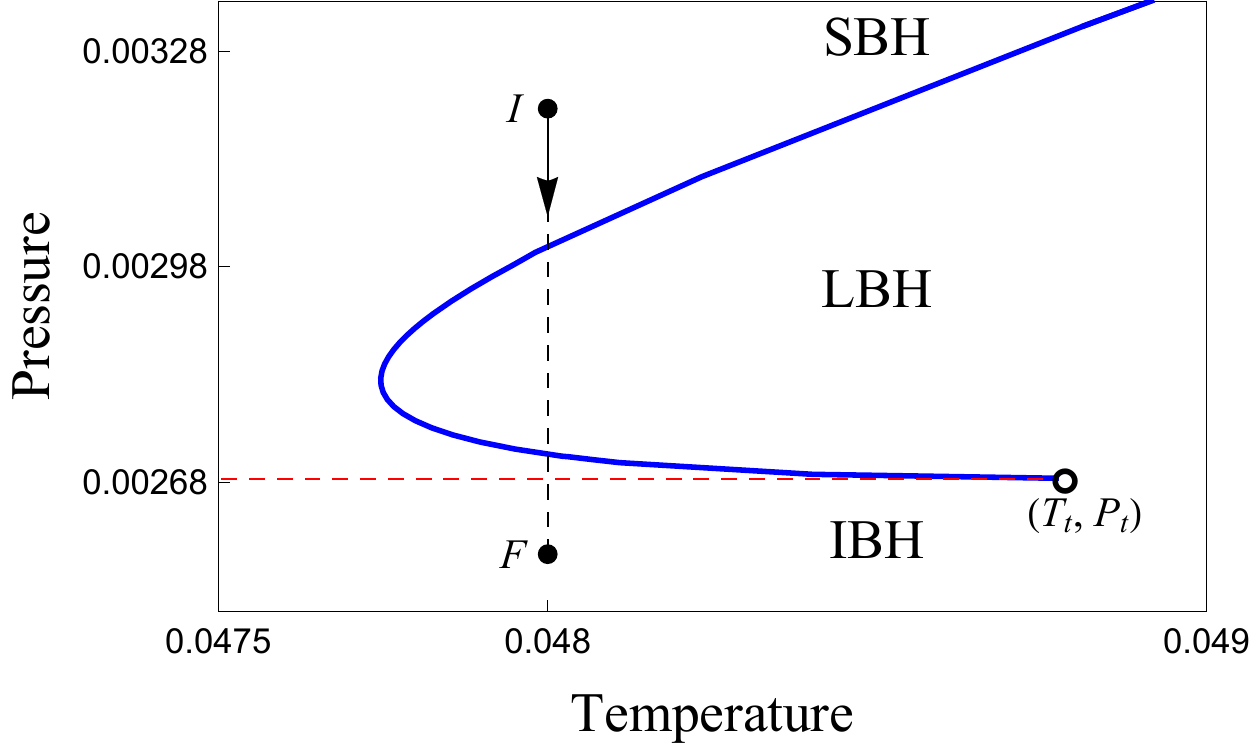}
    \caption{{\textbf{Reentrant phase transition.} The inset of Fig.~\ref{fig: RPT}b displays the pressure driven reentrant phase transition present for $\mu=0.15$. Namely, as we follow the black dashed curve from the initial point $I$ of high pressure, to the final point $F$ of low pressure, we cross twice the first order coexistence line (blue curve) and once the zeroth order coexistence line (denoted by dashed red curve), undergoing a reentrant phase transition from SBH to LBH back to SBH and finally to IBH. This is a first example where the reentrance is observed for black holes with the pressure as a control parameter.  }}
    \label{fig: zeroth2}
\end{figure}

As the string tension $\mu$ increases, a region of reentrant phase transitions emerges, illustrated in Fig.~\ref{fig: RPT}b for $\mu=0.15$. For values of $\mu$ in this intermediate range, the co-existence line of the first order phase transitions becomes double-valued, indicating  the presence of a pressure driven {\em reentrant phase transition}, shown in greater detail in Fig.~\ref{fig: zeroth2}.  As we follow the black dashed curve from the initial point $I$ of high pressure, to the final point $F$ of low pressure, we cross twice the first order coexistence line (blue curve) and once the zeroth order coexistence line (denoted by dashed red curve), undergoing a reentrant phase transition from SBH to LBH back to SBH and finally to IBH. This is the first example where reentrance is observed for black holes with the pressure as the control parameter.

Finally, for sufficiently large tensions, $\mu\approx 1/4$, the slope of the whole coexistence line becomes negative, as pressure decreases from $P_c$, temperature increases instead of decreases along the line, see Fig.~\ref{fig: RPT}c.  This
transition seems very different from that of the  liquid/gas phase transition.

\section{Summary and conclusion}
\label{sec: conclusion}

We have studied the phase behaviour of slowly accelerating charged AdS black holes, employing the recently derived thermodynamic quantities  \cite{Anabalon:2018qfv}.  We have found that whereas the high pressure behaviour resembles that of the non-accelerating case, there exists a second critical point  $P=P_t$ at low pressures where the swallowtail snaps. 
We have provided an intuitive explanation of this by looking at how the swallowtail plots cut the parameter space.  We can understand the snapping of the swallowtail as a result of
the existence of a critical slice through the point $X$ in the parameter space, together with the existence of the slow acceleration boundary.

The coexistence line of the first order phase transition thus terminates at $P_t$, where it is joined to a zeroth order coexistence line, across which we find a small to intermediate black hole phase transition. We  refer to the point  where the two coexistence lines intersect as a bicritical point, and denote it by $(P_t,T_t)$.

So far in our investigations we have concentrated on the case when only one (south pole) string is present and the north pole axis is regular, in which case the critical pressure $P_t$ is given
by formula \eqref{Pt1}. If both strings were present, the behaviour we have discovered would remain qualitatively the same, with the critical pressure instead given by 
\be\label{PtTT1}
P_t=\frac{3(\mu_+-\mu_-)^2}{8\pi Q^2}\,.
\ee
Since it is the difference of the two string tensions that causes the black hole to accelerate, we immediately see that the existence of $P_t$ is immediately linked to the black hole acceleration.

We also found a new form of a  reentrant phase transition from the double-valued co-existence curve, as well as a no black hole region in the $P-T$ plane.  This is the first example of a black hole reentrant phase transition in which
pressure is the parameter that montonically changes as we shift from one phase to another and then back to the first.

 Finally, we have discovered the existence of  (charged) mini-entropic black holes, whose temperature, electrostatic potential, and volume become  unboundedly large whilst their entropy, horizon size, and mass remain finite, with the latter approaching zero. The physical properties of these objects remain to be understood.

 We close by noting that there are two distinct viewpoints one can take concerning the first law \cite{pope2015}.
One is the  ``physical states" viewpoint, which regards the first law as describing  actual tiny physical perturbations to a given physical black hole, which subsequently settles down to another physical black hole described by parameters that are small changes from the original ones.  A more general ``equilibrium states" viewpoint
is that the first law describes  variations of the parameters characterizing a family of black hole solutions of interest.
Insofar as we have restricted our attention to slowly accelerating black holes, our results can at least be interpreted in the context of the latter viewpoint.  It remains an interesting question as to if and how the ``physical states" viewpoint can be adopted for  accelerating black holes  and what the newly obtained phase transitions   imply for the behaviour of the dual CFT.

 We have here considered the slow and fast cases  as (thermodynamically) `disconnected', neglecting the branch of fast accelerating black holes completely. Despite the fact that
fast accelerating black holes are `smoothly connected' to slowly accelerating ones in the parameter space, the asymptotic structures of the spacetimes they describe are completely different. A transition from slow to fast acceleration is in many respects similar to a transition from AdS to dS asymptotics:  new horizons appear in the fast accelerating case and change drastically the asymptotic and thermodynamic properties of the spacetime,  as a comparison of the corresponding Penrose diagrams \cite{Podolsky:2003gm}
clearly indicates.  Whether or not phase transitions can take place between them remains a question for future investigation.

Rotating and charged rotating black holes with non-zero acceleration are likely to yield even more interesting behaviour.  Work on investigating these cases is in progress.

\section*{Acknowledgments}
\label{sc:acknowledgements}

This work was supported in part by the Natural Sciences and Engineering Research Council of Canada.
D.K.\ acknowledges the Perimeter Institute for Theoretical Physics  for their support. Research at Perimeter Institute is supported by the Government of Canada through the Department of Innovation, Science and Economic Development Canada and by the Province of Ontario through the Ministry of Research, Innovation and Science.

\section*{References}
\bibliography{References}

\end{document}